\documentclass[dvips,12pt]{iopart}
\usepackage{graphics}
\usepackage{graphicx}
\usepackage{harvard}

\newcommand{\be}{\begin{equation}}
\newcommand{\ee}{\end{equation}}
\newcommand{\bea}{\begin{eqnarray}}
\newcommand{\eea}{\end{eqnarray}}
\newcommand{\figwidth}{0.7\linewidth}

\begin{document}

\note[Open-source software for generating 
electrocardiogram signals]{Open-source software for generating 
electrocardiogram signals}

\author{Patrick E. McSharry\dag\ddag\S\  and Gari D. Clifford$\Vert$\ }

\address{\dag\ Department of Engineering Science, University of Oxford, Parks Road, 
Oxford OX1 3PJ, UK}
\address{\ddag\ Mathematical Institute, University of Oxford, 24-29 St Giles', 
Oxford OX1 3LB, UK}
\address{\S\ Centre for the Analysis of Time Series, London School of Economics, 
London WC2A 2AE, UK}
\address{$\Vert$\ Harvard-MIT Division of Health Sciences \& Technology,
Rm E25-505, 45 Carleton St., Cambridge MA 02142, US}
\eads{\mailto{mcsharry@robots.ox.ac.uk}, \mailto{gari@mit.edu}}

\begin{abstract}
ECGSYN, a dynamical model that faithfully reproduces the main features of
the human electrocardiogram (ECG), including heart rate variability,
RR intervals and QT intervals is presented. Details of the underlying 
algorithm and an open-source software implementation in Matlab, C and Java 
are described.  An example of how this model will facilitate comparisons 
of signal processing techniques is provided.
\end{abstract}


\submitto{\PMB} 

\maketitle

\section{Introduction} 

The field of biomedical signal processing has given rise to a number of 
techniques for assisting physicians with their 
everyday tasks of diagnosing and monitoring medical disorders.  
Analysis of the electrocardiogram (ECG) provides a quantitative 
description of the heart's electrical activity and is routinely used in 
hospitals as a tool for identifying cardiac disorders.  

A large variety of signal processing techniques have been employed for filtering
the raw ECG signal prior to feature extraction and diagnosis of medical disorders.
A typical ECG is invariably corrupted by (i) electrical interference from surrounding 
equipment ({\it e.g.} effect of the electrical mains supply), (ii) measurement (or 
electrode contact) noise, (iii) electromyographic (muscle contraction), 
(iv) movement artefacts,  (v) baseline drift and respiratory artefacts  and 
(vi) instrumentation noise (such as artefacts from the analogue to digital 
conversion process) \cite{friesen90}. 

Many techniques may be employed for filtering 
and removing noise from the raw ECG signal, such as wavelet decomposition \cite{nikolaev00},
Principal Component Analysis (PCA) \cite{paul00}, Independent 
Component Analysis (ICA) \cite{potter02}, nonlinear noise 
reduction \cite{schreiberk96} and traditional Wiener methods.
The ECG forms the basis of a wide range of medical studies, including the 
investigation of heart rate variability, respiration and QT dispersion \cite{malik95}.  
The utility of these medical indicators relies on signal processing techniques for 
detecting R-peaks \cite{pan85}, deriving heart rate and respiratory rate \cite{moody85}, 
and measuring QT-intervals \cite{davey99}.  

Despite the numerous techniques that may be found in the literature and those that are 
now freely available on the Internet \cite{physionet}, it remains extremely difficult to 
evaluate and contrast their performance.  
The recent proliferation of biomedical databases, such as {\it Physiobank} \cite{physionet},    
provides a common setting for comparing techniques and approaches.  While this 
availability of real biomedical recordings has and will continue to advance the pace of 
research, the lack of internationally agreed upon benchmarks means that it is impossible 
to compare competing signal processing techniques.  The definition of such benchmarks 
is hindered by the fact that the true underlying dynamics of a real ECG can never be known.  
This void in the field of biomedical research calls for a {\it gold standard}, where an ECG 
with well-understood dynamics and known characteristics is made freely available. 

The model presented here, known as ECGSYN (synthetic electrocardiogram), is motivated by 
the need to evaluate and quantify the performance of the above signal processing techniques 
on ECG signals with known characteristics.
While the {\it Physionet} web-site \cite{physionet} already contains a synthetic ECG 
generator \cite{ruha97}, this is not intended to be highly realistic.  
The model and its underlying algorithm described in detail in this paper is capable 
of producing extremely realistic ECG signals with complete flexibility over the choice of 
parameters that govern the structure of these ECG signals in the temporal and spectral domains. 
In addition, the average morphology of the ECG may be specified.  
In order to facilitate the use of ECGSYN, software has been made freely 
available as both Matlab and C code \footnote{www.physionet.org/physiotools/ecgsyn}. 
Furthermore, users can employ ECGSYN over the Internet using a Java applet, which 
provides a means of downloading an ECG signal with characteristics selected from a 
graphical user interface.

\section{Background}
\begin{figure}
\begin{center}
\includegraphics[width=\figwidth]{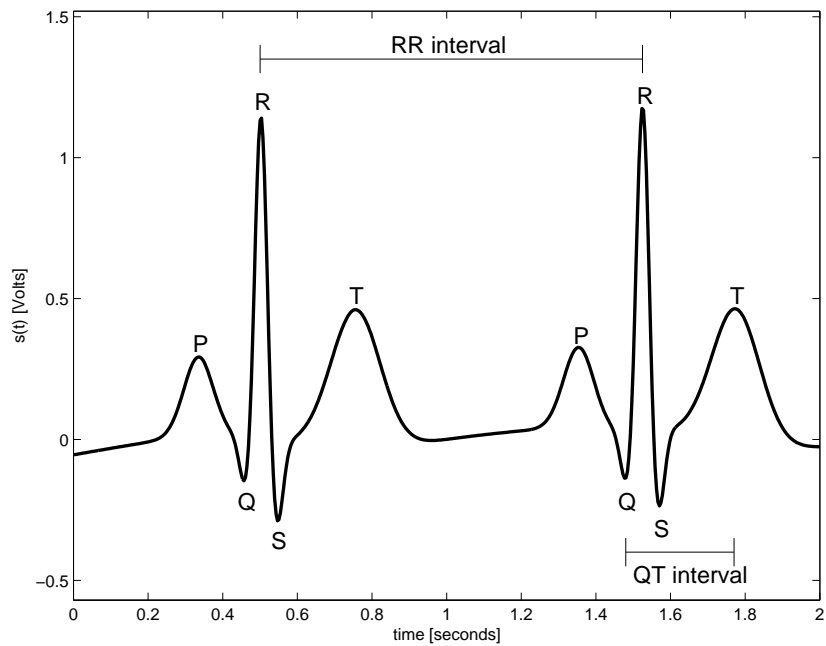}
\end{center}
\caption{Two seconds of synthetic ECG reflecting the electrical activity in the heart during 
two beats.  Morphology is shown by five extrema P,Q,R,S and T.  Time intervals corresponding to 
the RR interval and the QT interval are also indicated.}
\label{f:ecgmorph}
\end{figure}

The average heart rate is calculated by first measuring the time interval, denoted 
RR interval, between two consecutive R peaks (Fig. \ref{f:ecgmorph}), taking the 
average reciprocal of this value over a fixed window (usually 15, 30 or 60 seconds) 
and then scaling to units of beats per minute (bpm). 
A time series of RR intervals is often referred to as an RR tachogram and the variation in this
time series is governed by the balance between the sympathetic ({\it fight and flight}) and 
parasympathetic ({\it rest and digest}) branches of the central nervous system, known as the
sympathovagal balance. In general, innervation of the fast acting parasympathetic branch 
decreases heart rate, whereas the (more slowly acting) sympathetic branch increases heart rate.
This RR tachogram can therefore be used to estimate the effect of both these branches. 
A spectral analysis of the RR tachogram is usually divided into main frequency bands, 
known as the low-frequency (LF) band (0.04 to 0.15 Hz) and high-frequency (HF) band 
(0.15 to 0.4 Hz) \cite{eursoccard96}. Sympathetic tone is believed to affect the LF component 
whereas both sympathetic and parasympathetic activity influence the HF component 
\cite{malik95}. 
The ratio of the power contained in the LF and HF components has been used as a measure of the 
sympathovagal balance \cite{malik95,eursoccard96}. 

The structure of  the power spectrum of the RR tachogram tends to vary from person to person 
with a number of spectral peaks associated with particular biological 
mechanisms \cite{mcsharrycts02b,stefanovska02}. 
While the correspondence between these mechanisms and the positions of spectral peaks are 
strongly debated, there are two peaks which usually appear in most subjects. 
These are due to Respiratory Sinus Arrhythmia (RSA) \cite{hales1733,ludwig1847} caused by 
parasympathetic activity which is synchronous with the respiratory cycle and 
Mayer waves caused by oscillations in the blood pressure waves \cite{deboer87}.  
RSA usually gives rise to a peak in the HF region around 0.25 Hz, corresponding to 15 breaths 
per minute, whereas the Mayer waves cause a peak around 0.1 Hz.

\section{Method}

\begin{figure}
\begin{center}
\includegraphics[width=\figwidth]{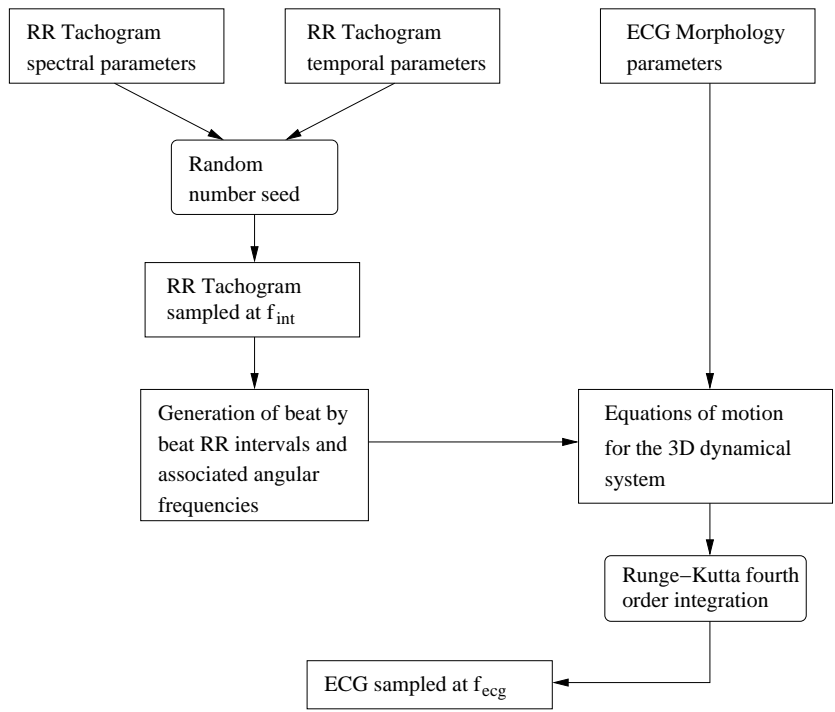}
\end{center}
\caption{\label{f:ecgsynflow}
ECGSYN flow chart describing the procedure for specifying the temporal and spectral 
description of the RR tachogram and ECG morphology.}
\end{figure}

The dynamical model, ECGSYN, employed for generating the synthetic ECG is composed of two 
parts.  
Firstly, an internal time series with internal sampling frequency  
$f_{\rm int}$ is produced to incorporate a specific mean heart rate, standard 
deviation and spectral characteristics corresponding to a real RR tachogram. 
Secondly, the average morphology of the ECG is produced by specifying the locations and heights 
of the peaks that occur during each heart beat.  A flow chart of the processes 
in ECGSYN for producing the ECG is shown in Fig. \ref{f:ecgsynflow}. 

The spectral characteristics of the RR tachogram, including both RSA and Mayer waves, 
are replicated by specifying a bi-modal spectrum composed of the sum of two Gaussian functions, 
\be
S(f) = \frac{\sigma_1^2}{\sqrt{2 \pi c_1^2}}
\exp \left( \frac{(f - f_1)^2}{2 c_1^2} \right)
+ \frac{\sigma_2^2}{\sqrt{2 \pi c_2^2}}
\exp \left( \frac{(f - f_2)^2}{2 c_2^2} \right),
\label{e:Sf}
\ee
with means $f_1,f_2$ and standard deviations $c_1,c_2$.  
Power in the LF and HF bands are given by
$\sigma_1^2$ and $\sigma_2^2$ respectively whereas the variance
equals the total area $\sigma^2 = \sigma^2_1+\sigma^2_2$,
yielding an LF/HF ratio of $\sigma^2_1/\sigma^2_2$.

A time series $T(t)$ with power spectrum $S(f)$ is generated by
taking the inverse Fourier transform of a sequence of complex numbers with
amplitudes $\sqrt{S(f)}$ and phases which are randomly
distributed between 0 and $2 \pi$.
By multiplying this time series by an appropriate
scaling constant and adding an offset value, the resulting time series can be
given any required mean and standard deviation. 
Different realisations of the random phases may be specified by varying the seed 
of the random number generator. In this way, many different time series $T(t)$ may 
be generated with the same temporal and spectral properties.

\begin{figure}
\begin{center}
\includegraphics[width=\figwidth]{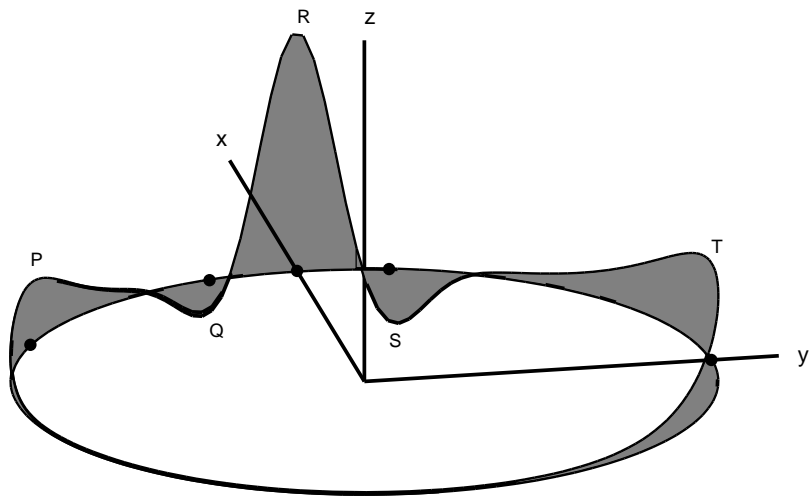}
\end{center}
\caption{\label{f:ecgmorph3d}
Three-dimensional state space of the dynamical system given by (\ref{e:pqrst}) 
showing motion around the limit cycle in the horizontal $(x,y)$-plane.
The vertical $z$-component provides the synthetic ECG signal with a morphology that is 
defined by the five extrema P,Q,R,S and T.}
\end{figure}

The ECG traces a quasi-periodic waveform with each beat of the heart, with the morphology 
of each cycle labeled by its peaks and troughs, P, Q, R, S and T, as shown in 
Fig. \ref{f:ecgmorph}.  
This quasi-periodicity can be reproduced by constructing a dynamical model containing an 
attracting limit cycle;  each heart beat corresponds to one revolution around this  
limit cycle, which lies in the $(x,y)$-plane as shown in Fig. \ref{f:ecgmorph3d}.  
The morphology of the ECG is created by using a series of exponentials to force the 
trajectory to trace out the PQRST-waveform in the $z$-direction.   A series of five angles, 
($\theta_P$, $\theta_Q$, $\theta_R$, $\theta_S$, $\theta_T$), are used to specify the 
extrema of the peaks (P,Q,R,S,T) respectively.  

The dynamical equations of motion are given by three ordinary
differential equations \cite{mcsharrycts03url},  
\bea
{\dot x} &=& \alpha x  - \omega y, \nonumber \\
{\dot y} &=& \alpha y + \omega x, \nonumber \\
{\dot z} &=& - \!\!\!\!\!\! \sum_{i \in \{P,Q,R,S,T\}} \!\!\!\!\!\!
a_i \Delta \theta_i
\exp(-\Delta \theta_i^2 / 2 b_i^2) -  (z - z_0),
\label{e:pqrst}
\eea
where $\alpha = 1 - \sqrt{x^2 + y^2}$,
$\Delta \theta_i = (\theta - \theta_i) \ {\rm mod} \ 2 \pi$,
$\theta = {\rm atan2}(y,x)$ and $\omega$ is the angular velocity
of the trajectory as it moves around the limit cycle.
The coefficients $a_i$ govern the magnitude of the peaks whereas the $b_i$ define the width 
(time duration) of each peak.  
Baseline wander may be introduced by coupling the baseline value $z_0$
in (\ref{e:pqrst}) to the respiratory frequency $f_2$ in (\ref{e:Sf}) using
$z_0(t) = A \sin(2 \pi f_2 t)$.  The output synthetic ECG signal, $s(t)$, is the 
vertical component of the three-dimensional dynamical system in (\ref{e:pqrst}): $s(t) = z(t)$.

Having calculated the internal RR tachogram expressed by the time series $T(t)$ with power 
spectrum $S(f)$ given by (\ref{e:Sf}), this can then be used to drive the dynamical 
model (\ref{e:pqrst}) so that the resulting 
RR intervals will have the same power spectrum as that given by $S(f)$.  
Starting from the auxiliary\footnote{This auxiliary time axis is used to calculate the 
values of $\Omega_n$ for consecutive RR intervals whereas the time axis for the ECG 
signal is 
sampled around the limit cycle in the $(x,y)$-plane.}
time $t_n$, with angle $\theta = \theta_R$, the time interval $T(t_n)$ 
is used to calculate an angular frequency $\Omega_n = \frac{2 \pi}{T(t_n)}$.  
This particular angular frequency, $\Omega_n$, is used to specify the dynamics 
until the angle $\theta$ reaches $\theta_R$ again, whereby a complete revolution 
(one heart beat) has taken place.  For the next revolution, 
the time is updated, $t_{n+1} = t_n + T(t_n)$, and the next angular frequency, 
$\Omega_{n+1} =  \frac{2 \pi}{T(t_{n+1})}$, is used to drive the trajectory around 
the limit cycle. In this way, the internally generated beat-to-beat 
time series, $T(t)$, can be used to generate an ECG with associated RR intervals 
that have the same spectral characteristics. 
The angular frequency $\omega(t)$ in (\ref{e:pqrst}) is specified using the 
beat-to-beat values $\Omega_n$ obtained from the internally generated RR tachogram:
\be
\omega(t) = \Omega_n, \ \ \ \ \ \ t_n \leq t < t_{n+1}.
\ee

Given these beat-to-beat values of the angular frequency $\omega$, the equations of 
motion in (\ref{e:pqrst}) are integrated using a fourth-order Runge-Kutta 
method \cite{press92}. 
The time series $T(t)$ used for defining the values of $\Omega_n$ has a high sampling 
frequency of  $f_{\rm int}$, which is effectively the step size of the integration. 
The final output ECG signal is then down-sampled to $f_{\rm ecg}$
if $f_{\rm int} > f_{\rm ecg}$ by a factor $\frac{f_{\rm int}}{f_{\rm ecg}}$
to generate an ECG at the requested sampling frequency.
For simplicity, $f_{\rm int}$ is taken as an integer multiple of $f_{\rm ecg}$ and  
anti-aliasing filtering is therefore not required if $f_{\rm ecg}$ is chosen to be 
sufficiently high.

\begin{figure}
\begin{center}
\includegraphics[width=\figwidth]{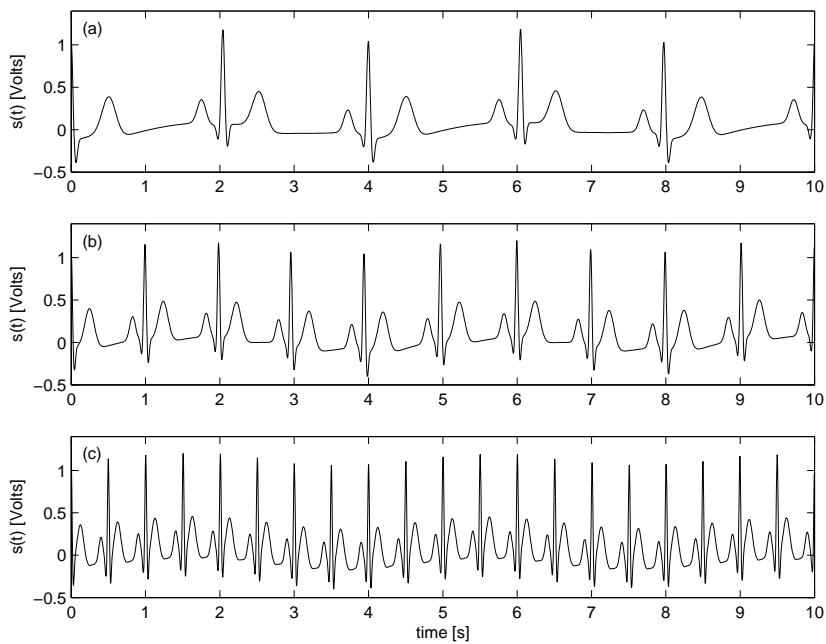}
\end{center}
\caption{\label{f:ecgheartrates} 
Synthetic ECG signals for different mean heart rates: (a) 30 bpm, (b) 60 bpm and 
(c) 120 bpm.}
\end{figure}

The size of the mean heart rate affects the shape of the ECG morphology.   
An analysis of real ECG signals for different heart rates shows that the intervals 
between the extrema vary by different amounts; in particular the QRS width 
decreases with increasing heart rate. This is as one would expect;
when sympathetic tone increases the conduction velocity across the ventricles 
increases, together with an augmented heart rate. The time for ventricular 
depolarisation (represented by the QRS complex of the ECG)  is therefore 
shorter. 
These changes are replicated by modifying the width of the exponentials in (\ref{e:pqrst}) 
and also the positions of the angles $\theta$. This is achieved by using a heart rate 
dependent factor $\alpha = \sqrt{h_{\rm mean}/60}$ where $h_{\rm mean}$ is the mean heart 
rate expressed in units of bpm (see Table \ref{t:pqrst}).

Operation of ECGSYN, composed of the spectral characteristics given by (\ref{e:Sf}) and 
the time domain dynamics in (\ref{e:pqrst}), requires the selection of the 
list of parameters given in Tables \ref{t:pqrst} and \ref{t:params}.

\begin{table}
\caption{\label{t:pqrst} Morphological parameters of the ECG model with modulation factor 
$\alpha = \sqrt{h_{\rm mean}/60}$.}
\begin{indented}
\item[]\begin{tabular}{llllll}
\br
Index (i) &P    &Q    &R    &S   &T \\
\mr 
Time (secs) &-0.2 &-0.05 &0  &0.05 &0.3 \\
$\theta_i$ (radians) &$-\frac{1}{3}\pi\sqrt{\alpha}$ &$-\frac{1}{12}\pi\alpha$
&0 &$\frac{1}{12}\pi\alpha$ &$\frac{1}{2}\pi$  \\
$a_i$ &1.2  &-5.0  &30.0 &-7.5 &0.75 \\
$b_i$ &0.25$\alpha$  &0.1$\alpha$  &0.1$\alpha$ &0.1$\alpha$ &0.4$\alpha$ \\
\br
\end{tabular}
\end{indented}
\end{table}

\begin{table}
\caption{\label{t:params} Temporal and spectral parameters of the ECG model}
\begin{indented}
\item[]\begin{tabular}{lll}
\br
Description                          &Notation       &Default values \\ 
\mr 
Approximate number of heart beats    &$N$            &256     \\
ECG sampling frequency               &$f_{\rm ecg}$  &256 Hz  \\
Internal sampling frequency          &$f_{\rm int}$  &512 Hz  \\
Amplitude of additive uniform noise  &$A$            &0.1 mV  \\
Heart rate mean                      &$h_{\rm mean}$ &60 bpm  \\
Heart rate standard deviation.       &$h_{\rm std}$  &1  bpm  \\
Low frequency                        &$f_1$          &0.1 Hz  \\
High frequency                       &$f_2$          &0.25 Hz \\ 
Low frequency standard deviation     &$c_1$          &0.1 Hz  \\
High frequency standard deviation    &$c_2$          &0.1 Hz  \\
LF/HF ratio                          &$\gamma$       &0.5     \\ 
\br 
\end{tabular}
\end{indented}
\end{table}

\section{Results}

The synthetic ECG provides a realistic signal for a range of heart rates.   
Figure \ref{f:ecgheartrates} illustrates examples of the synthetic ECG for three 
different heart rates; 30 bpm, 60 bpm, and 120 bpm. Notice that the PR, QT and QRS widths 
all shorten with increasing heart rate. It is important to note that the nonlinear 
relationship between the morphology modulation factor $\alpha$ and mean heart rate 
$h_{\rm mean}$ limits the contraction of the overall PQRST morphology relative to the 
refractory period (the minimum amount of time in which depolarisation and repolarisation 
of the cardiac muscle can occur).

\begin{figure}
\begin{center}
\includegraphics[width=\figwidth]{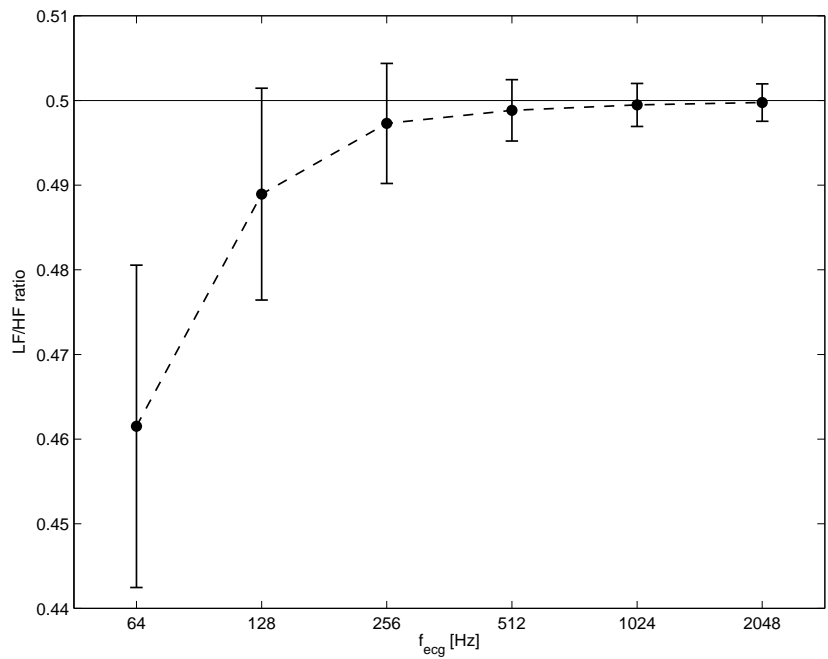}
\end{center}
\caption{\label{f:lfhfsf}
LF/HF ratio estimates computed from synthetic ECG signals for a range of sampling 
frequencies using an input LF/HF ratio of 0.5 (horizontal line).  The distribution of 
estimates is shown by the mean (dot) and plus/minus one standard deviation error bars.  
The simulations used 100 realisations of noise-free synthetic ECG signals with a mean heart 
rate of 60 bpm and standard deviation of 3 bpm.}
\end{figure}

The ability of ECGSYN to generate ECG signals with known spectral characteristics provides 
a means of testing the effect of 
varying the ECG sampling frequency $f_{\rm ecg}$
on the estimation of heart rate variability (HRV) metrics.  
Figure \ref{f:lfhfsf} illustrates the increase in estimation accuracy of a HRV metric, 
the LF/HF ratio, with increasing $f_{\rm ecg}$. 
The error bars represent one standard deviation on either 
side of the means (dots) of each 1000 Montecarlo runs. 
The true input LF/HF ratio was 0.5 as shown by the horizontal line.  
The synthetic ECG signals had a mean heart rate of 60 bpm and a standard deviation of 3 bpm.  
The method used for estimating the LF/HF ratio, the Lomb periodogram, introduces negligible 
variance into the estimate \cite{clifforddphil}, and therefore the downward bias of the 
estimates can be considered due to 
$f_{\rm ecg}$ being too low.
Note that below 512 Hz, the LF/HF ratio is considerably underestimated. 
This is consistent with studies performed on real data \cite{abboud95}.

\section{Discussion}

A dynamical model known as ECGSYN has been presented that generates realistic synthetic ECG 
signals.  The user can specify both the temporal and spectral characteristics of the ECG.  
In addition, the 
average morphology of the ECG may be input into the algorithm.  
Open-source software for the algorithm underlying ECGSYN 
is freely available in both Matlab and C.  A Java applet facilitates the 
generation of ECG signals over the Internet with characteristics selected using a graphical 
user interface.    

By examining the statistical properties of artificially generated ECG signals, it has been 
shown that estimates of HRV using the LF/HF ratio depend on the sampling frequency, 
$f_{\rm ecg}$, of the ECG. Small values of $f_{\rm ecg}$ gives rise to ECG signals which lead 
to underestimated LF/HF ratios. 
This provides a basis for the low sample frequency problem in HRV 
studies \cite{abboud95}.  In addition, these results provide a guide for physicians when 
selecting the sampling frequency of the ECG based on the required accuracy of the HRV metrics.  

The availability of ECGSYN through open-source software 
and the ability to generate collections of ECG signals with carefully 
controlled and {\it a priori} known characteristics will allow biomedical researchers to 
test and provide operation statistics for new signal processing techniques.  
This will enable physicians to compare and evaluate different techniques and to select 
those that best suit their requirements.

\ack 

PEM acknowledges support of a Research Fellowship 
from the Royal Academy of Engineering and the Engineering and Physical Sciences 
Research Council (EPSRC). GDC acknowledges support by the US National Institute of 
Health (NIH), grant number EC001659-01.  The authors would like to thank Mauricio Villarroel 
for developing the Java applet for ECGSYN.

\end{document}